# NON-INVASIVE DEEP-BRAIN STIMULATIONS BY SPATIO-TEMPORAL FOURIER SYNTHESIS [‡]

Laszlo B. Kish [§,1], Andrea Antal [2]

[1] *Department of Electrical and Computer Engineering, Texas A&M University, College Station, Texas, USA*
[2] *Department of Neurology, Robert Koch Str. 40, D-57075 Göttingen, Germany*

*Laszlokish@tamu.edu , aantal@gwdg.de*

A new type of non-invasive deep-brain stimulation is conceived and demonstrated by computer simulations. The process is based on spatio-temporal Fourier synthesis using multiple electrode pairs with sinusoidal current drive to limit skin sensations and concentrate the stimulus power to a small spatial volume and into large rare spikes in the times domain, while the signal power at the skin is steady and small. Exotic time signals are also shown, such as the cases of high-frequency prime harmonics, quasi-random and chirping stimulations. The first one is able to generate sharp spikes with low frequency while its carriers are high-frequency harmonics that easily conducts via the skin and brain tissue. Open questions are, among others, the best shapes and timing of spikes. The answers require experimental tests and explorations in animal models and human subjects.

## I. Introduction: Deep-Brain Stimulations with Multiple Electrode Pairs

Deep brain stimulation (DBS) [1] is old field of neuroscience and they offer the potential to discover new brain targets for treating neurological and psychiatric disorders like Parkinson's disease, chronic pain, depression, and obsessive-compulsive disorder. However, the old way to implant electrodes into the brain is invasive and it has high risks involved thus non-invasive DBS techniques [2] have been developed. Among others, promising alternatives had been transcranial magnetic and low intensity electrical stimulations (TMS, tES where the small "t" stands for low intensity) however they do not reach deep enough due to the power-function decay of the dipole field and they excite mostly the shallow cortical layers.

*1.1 On non-invasive, electrical deep-brain stimulations*

In the present paper, stimulations based on transcranial currents are discussed. These techniques (e.g. [2-9]) use external electrodes placed on the scalp to deliver electrical stimulation deep within the brain without the need for surgery. The current flows through the bulk of the brain and the ultimate questions are: How to spatially focus the stimulus on the critical subvolume in the brain, and how to temporally reach the necessary levels of stimulus with minimal side and adverse effects.

*1.2 Transcutaneous electrical stimulations with currents of beating frequencies*

One of the recent techniques (e.g. [5-9]) is called temporal interference (TI) stimulation. This non-invasive deep-brain stimulation has first been proposed in [5]. The technique uses electrode pairs driven by "high-frequency" (kiloHertz range, which counts as high-frequency in the field of brain signals) sinusoidal signals with slightly differing frequencies. This approach results in a low-frequency beat of the excitation power in the brain area where these current paths intersect.

---

[‡] Texas A&M University patent disclosure was submitted on January 15, 2024. Provisional patent application was filed with U.S. Serial No.: 63/574,412. A slightly modified version will orally be presented at the 9th International Conference on Unsolved Problems of Noise, Budapest, June 3-7, 2024.
[§] Corresponding author. Honorary faculty at Óbuda University, Budapest, Bécsi út 96/B, Budapest, H-1034, Hungary.

Mathematically, the theory of amplitude modulation of AM radio stations is relevant, where a sinusoidal carrier with frequency $f_c$ is multiplied by a much lower signal frequency $f_s$, resulting in two additive sinusoidal signals with frequency difference $2f_s$:

$$\cos\left[2\pi t(f_c + f_s)\right] - \cos\left[2\pi t(f_c - f_s)\right] = 2\cos(2\pi f_c t)\cos(2\pi f_s t) . \qquad (1)$$

In the case of TI stimulation, the left hand side of Equation 1 represents the shape of the two high-frequency sinusoidal currents, and the right hand side is the resultant interference, which is a high-frequency sinusoidal current that is beating with the half of the difference of the frequencies. As an illustration, see Figure 1.

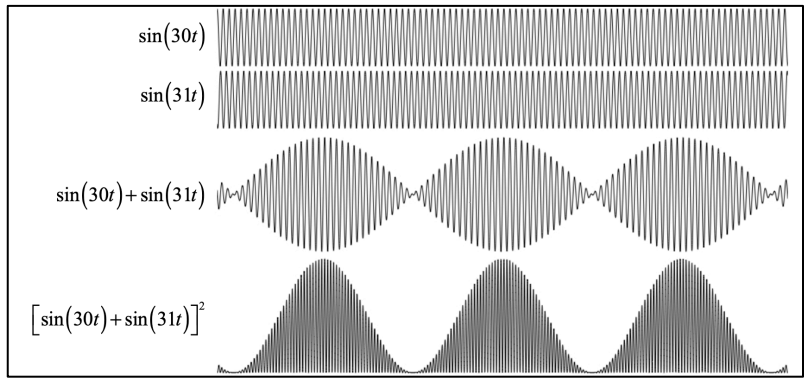

Figure 1. Illustration by Apple Grapher: The interference of currents sin(30t) and sin(31t) results in a beating envelope with the frequency of 0.5. The envelope of the square of the beating current represents the relevant excitation power, which is shown in the lowest graph.

The high-frequency (kiloHertz-range) currents are entering through two, non-overlapping electrode pairs at levels where the skin reactions are tolerable and in the "sweet-spot", where the current paths cross, the interfering current densities can theoretically at most double at the temporal peak. Although a high-frequency stimulation itself may not have a significant effect, the stimulus can still be enhanced due to nonlinear (rectification) effects. The nonlinear component with the lowest order in the Taylor polynomial is the second order. For an illustration of that effect, the square of the resulting beating current is also shown with up to four times enhancement compared to the case of a single high-frequency signal.

The only apparent disadvantage of the beating resultant current density is that its duty cycle is large. One can theorize that neurons would rather prefer small duty cycle, that is, excitations with short spikes, such as the neural signals are. On the other hand, one can also envision, that such large duty cycle excitations would be preferable to stimulate muscles due to the enhanced duration of the maxima. Experiments confirm these theoretical expectations, see below.

Researchers have shown that TI current beats are properly distributed in models and cadaver brains [6,7] and they are excellent stimulators of muscles in humans [8,9].

However, the actual stimulation of the brain of living humans has been challenging so far. While some studies have achieved a limited brain stimulation with short periods of special signal patterns, significant efforts have not always resulted in successful brain stimulation

[9].

The present work proposes a novel technique to achieve enhanced spatial confinement of the resultant stimulation current within the targeted neural tissue. Additionally, the method aims to temporally concentrate the current into brief, high-amplitude pulses for efficient neural activation using minimal single-electrode current amplitudes.

**2. The Proposed New Method for Deep-Brain Stimulations**

The method is not using beating sinusoidal currents. Our goal is to utilize Spatio-Temporal Fourier Synthesis (STFS) to focus the excitation both in space and time. The stimulation takes place via multiple electrode pairs driven by properly chosen sinusoidal harmonics of a periodic, quasi-periodic, or chirped Fourier series. Different harmonics feed the different electrode pairs. The more electrode pairs are the better as the upper limit of the peak excitation by the square of the stimulation scales with the second power of the number of electrode pairs.

*2.1 Principle of the spatial focus of the stimulus*

An illustration of a 4-electrode-pair system is shown in Figure 2. In this example, each electrode pair is driven by different ground-independent, AC current generators supplying proper sinusoidal harmonics $I_1(t)$, $I_2(t)$, $I_3(t)$, $I_4(t)$ of uniform effective value, that is:

$$I_1 = I_2 = I_3 = I_4 \quad , \tag{2}$$

where the currents without time coordinate represent RMS (effective) values.

The power is concentrated in the volume where the current density components originating from the different electrode pairs are equal. In this illustrative example, the skin-skull-brain system is modeled as a spherical geometry. The goal is to stimulate the center of the brain.

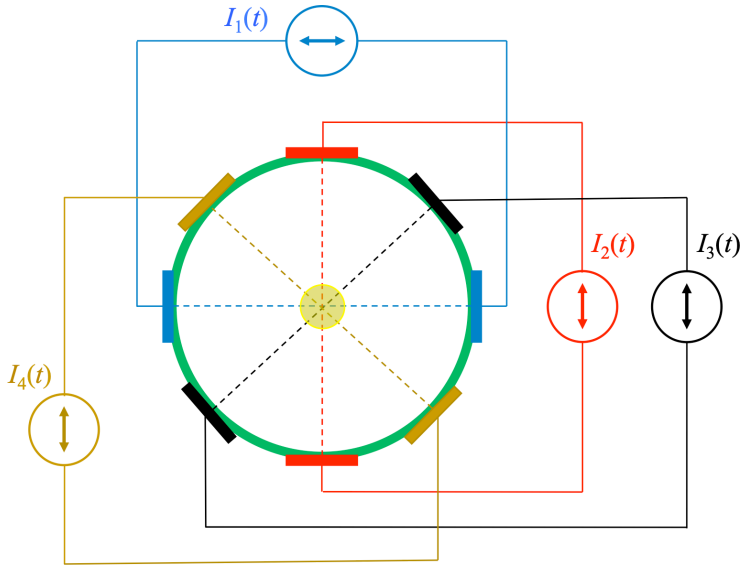

Figure 2. Example of the distributed stimulation with four electrode pairs. In this example, the currents are driven through the four AC current generators $I_1(t)$, $I_2(t)$, $I_3(t)$, $I_4(t)$, respectively, and they are superimposed in the deep-brain with equal weights. The goal is to stimulate the center part. For the sake of simplicity, the field lines have been simplified and homogeneity has been assumed for illustrative purposes.

The electrodes of the electrode pairs are placed diagonally. At the center, the AC current densities arising from the different electrode pairs have the same effective values due to symmetry reasons. This allows the application of the theory of generating large spikes of stimulation with small duty cycle.

If the goal is to stimulate a different region, the same electrode setup can be used, but the effective values of the current generators must be tuned such that the uniform harmonics situation occurs in the desired part of the brain.

*2.1 Theoretical scheme of the temporal focus of the stimulus*

The theory of an infinite Fourier series [10] indicate that a periodic spike train of uniform Dirac pulses in time results in a frequency spectrum of uniform lines at all the harmonics. This fact suggest ways about generating current density impulses of sharp spikes by the addition of sinusoidal current components. The resulting current density is shown below:

$$J(t) = J_1 \cos(2\pi f_1 t) + J_2 \cos(2\pi f_2 t) + ... + J_N \cos(2\pi f_N t) , \qquad (3)$$

where $J(t)$ is the total current density, $J_k$ (k=1, …, N) is the current density amplitude of the $k$-th component due to the electrode current $I_k(t)$, and $f_k$ is the corresponding frequency.

In general, we envision frequency harmonics however there are exceptions, such as the noise based stimulation where the frequencies are irrational numbers (quasi-periodic process), or when the frequencies are large prime numbers without existing first harmonics, see in Section 3.

Depending on the choice of the frequencies of the components, the $J(t)$ current density over the sweet spot can have various types of spiky behavior with stimulation focused not only in the spatial domain (Figure 2) but also in time.

In the following section, we will examine several promising avenues, though our exploration is not intended to be exhaustive.

## 3. Computer simulation examples of the temporal focus of the stimulation

The rectification that the ion channels are performing, is represented by the increase of the Taylor series components with even power exponent. The lowest of these is the exponent of 2. Therefore, to see the enhancement of the stimulus by nonlinearity/rectification effects, it is a pessimistic approach to plot the square of the resultant current densities.

*3.1 Sinusoidal current shapes*

In Figure 3, the following equation is plotted. The square of the sum of the first four cosinus harmonics is shown:

$$J^2(t) = \left[\cos(t) + \cos(2t) + \cos(3t) + \cos(4t)\right]^2 . \qquad (4)$$

The squares of the harmonics are also shown for comparison. The spikes enhance the stimulation effect by a factor of 16 and their repetition frequency is the equal to the frequency of the first harmonic.

The phase of the harmonics also matters. Using sinus functions in Equation 5 instead of the cosinus functions of Equation 4 causes major changes because the resulting phase shift is different for the different harmonics.

$$J^2(t) = \left[\sin(t) + \sin(2t) + \sin(3t) + \sin(4t)\right]^2 \quad . \tag{5}$$

The enhancement is only about 11, and the stimulation arises as twin spikes.

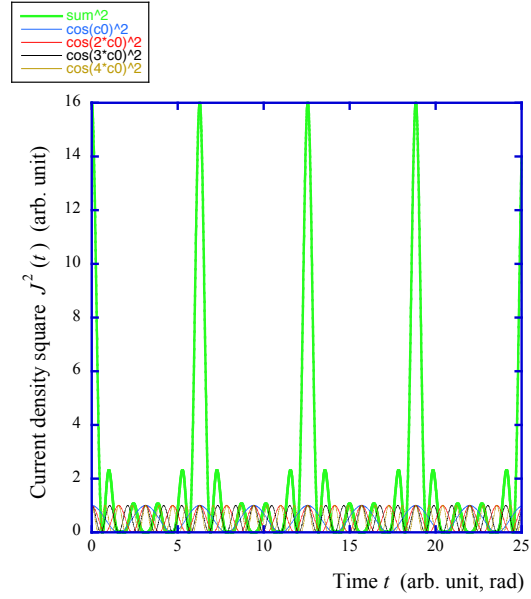

Figure 3. The square of the sum of the first four cosinus harmonics. The squares of the harmonics are also shown. The spikes enhance the stimulation effect by a factor of 16 and their repetition frequency is the equal to the frequency of the first harmonic.

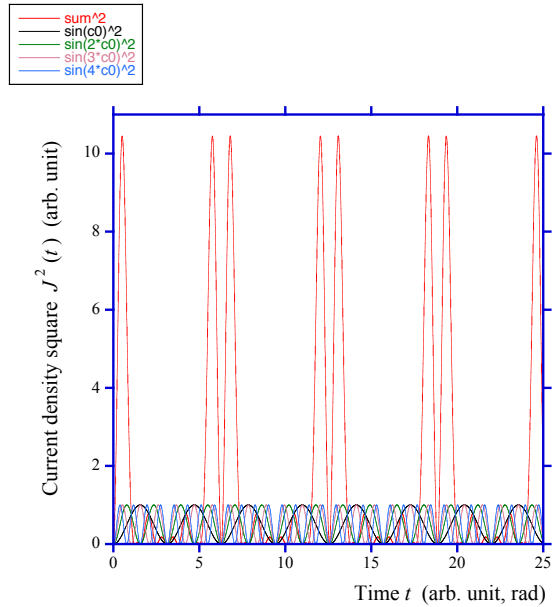

Figure 4. The phase of the harmonics does matter. Using sinus functions in Equation 5 instead of the cosine functions of Equation 4 causes major changes because the resulting phase shift is different for the different harmonics. The enhancement is only about 11, and the stimulation arises as twin spikes.

*3.2 Exotic current shapes: high-frequency prime-number-harmonics and chirping*

Changing the frequencies of the current components of the Fourier synthesis opens a rich set of new possibilities. We outline a few examples below.

### 3.2.1 High-frequency prime-number-harmonics for rare periodic spikes

If the goal is to carry the excitation through the skin via a high-frequency current carrier, then one of the possible solutions is to use high-frequency driving current harmonics with prime number frequencies. Then the square of the current density will show large short spikes repeating with the frequency of the first harmonics. In the example of Figure 5, the current components are 79, 83, 89 and 97 times over the frequency of the first harmonic. This situation results in about a 100 times higher average carrier frequency over the skin than the repetition frequency of the spikes.

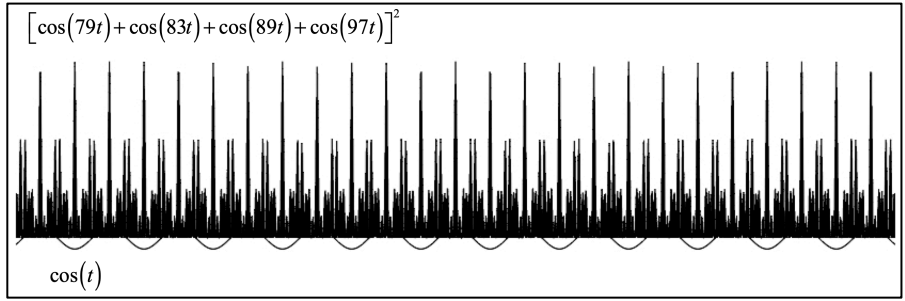

Figure 5. Example for four prime number harmonics without the first harmonics. The first harmonic is also plotted for time scale reference but only its negative maxima are visible. The carrier frequencies about hundred times (79, 83, 89, 97) higher.

### 3.2.2 Noisy excitations: Components with irrational frequency values

Example for noise stimulation by utilizing harmonics with irrational frequencies resulting in a quasi-periodic noise process is shown in Figure 6. It is an open question if noisy excitation is more effective.

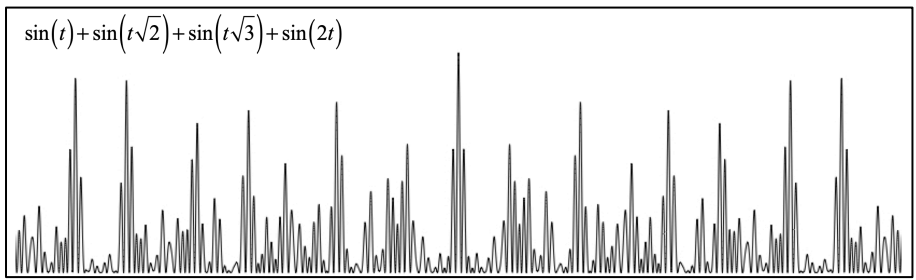

Figure 6. Noisy stimulation (quasi periodic noise). The frequencies of second and the third components are irrational numbers, while the first and the second harmonics are rational numbers.

### 3.2.3 Chirping frequencies

Figure 7 shows and example of chirping. In Equation 5 (and Figure 4), the frequencies are linearly growing in time. The resulting time function is interesting and it is an open question if it enhances stimulation or not.

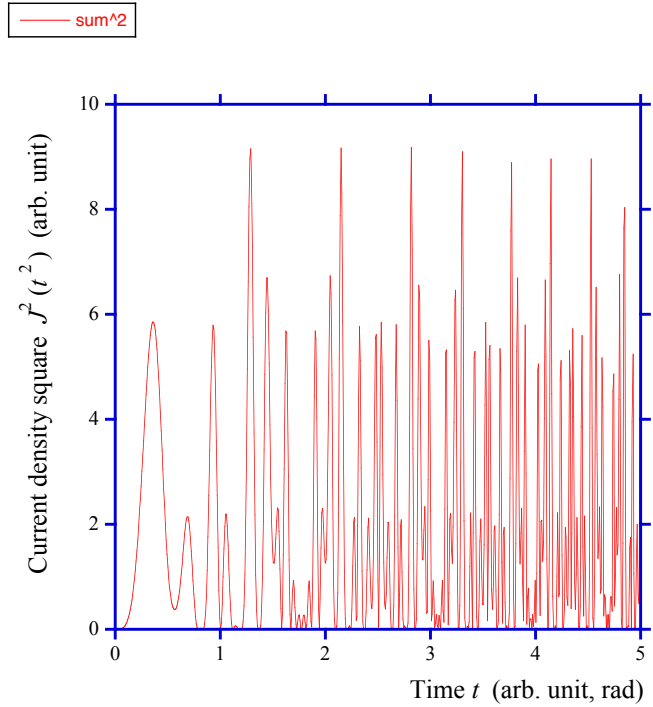

Figure 7. Chirping stimulus, with harmonic frequencies in Equation 5 and Figure 4 linearly increasing in time.

## 4. Conclusions

A novel non-invasive deep-brain stimulation technique has been introduced and demonstrated through computer simulations. The process utilizes spatio-temporal Fourier synthesis, employing multiple electrode pairs with sinusoidal current drive. This approach aims to limit skin sensation and concentrate the stimulus power within a small spatial volume, while generating large, intermittent spikes in the time domain. Conversely, the signal power at the skin remains steady and low.

The efficiency of the various stimulation methods described is an open question that requires further exploration. Key aspects to investigate include:

*Efficiency*: The overall efficiency of the stimulation techniques, particularly in terms of power delivery to the target deep brain regions.

*Skin Sensitivity and Tolerability*: The skin's sensitivity and the tolerability of the stimulation, which are crucial factors for patient comfort and acceptance.

*Electrode Pair Placement*: The optimal placement of the electrode pairs to achieve the desired deep brain targeting and stimulation.

*Waveforms*: An intriguing question is the identification of the most effective synthesized for these types of transcutaneous stimulations, as it may depend on the specific deep brain section targeted for impact.

An important question, how can we shift the location of the stimulation in the deep brain? There are three potential ways to do that:

(i) Modifying the locations of the electrode pairs, see [5] as example. The highest stimulation with the clear spiky waves described above arises in the brain subvolume where the RMS values of the four electrical fields are equal.

(ii) Modifying the RMS values of the driving currents of the different electrode pairs. Similarly as above, the clear spiky waves arise in the brain subvolume where the RMS values of the four electrical fields are equal.

(iii) Modifying type or the parameters of the wave signals (see the examples above). We hypothesize that different brain segments are particularly sensitive to specific wave types and parameters. Future will tell is this so.

Our paper aims to introduce a tool for a rich set of stimulation with spiky waves resembling neural spikes.

All these aspects will require experimental testing and validation to fully assess the potential of this non-invasive deep-brain stimulation approach.

**References**

[1] A. M. Lozano1, N. Lipsman, H. Bergmani, P. Browni, S. Chabardes, J. W. Chang, K. Matthews, C. C. McIntyre, T. E. Schlaepferi, M. Schulder, Y. Temel, J. Volkmann, J. K. Krauss, Deep brain stimulation: current challenges and future directions, *Nat. Rev. Neurol.* **15** (2019) 148–160; doi: 10.1038/s41582-018-0128-2

[2] X. Liu, F. Qiu, L. Hou, X. Wang, Review of Noninvasive or Minimally Invasive Deep Brain Stimulation, *Frontiers Behavioral Neurosci.* **15** (2022) 820017; doi: 10.3389/fnbeh.2021.820017

[3] A. Antal, C. S. Herrmann, Transcranial Alternating Current and Random Noise Stimulation: Possible Mechanism Neural Plasticity 2016 (2016) 3616807; doi: 10.1155/2016/3616807

[4] W. Potok, O. van der Groen, S. Sivachelvam, M. Bächinger, F. Fröhlich, L.B. Kish, N. Wenderoth, Contrast detection is enhanced by deterministic, high-frequency transcranial alternating current stimulation with triangle and sine waveform, *J. Neurophysiology* **130** (2023) 458–473. https://doi.org/10.1152/jn.00465.2022

[5] N. Grossman, David Bono, N. Dedic, L.-H. Tsai, A. Pascual-Leone, E. S. Boyden, Noninvasive Deep Brain Stimulation via Temporally Interfering Electric Fields, *Cell* **169** (2017) 1029–1041.

[6] I. R. Violante, K. Alania, A. M. Cassarà, E. Neufeld, E. Acerbo, R. Carron, A. Williamson, D. L. Kurtin, E. Rhodes, A. Hampshire, N. Kuster, E. S. Boyden, A. Pascual-Leone, N. Grossman, Non-invasive temporal interference electrical stimulation of the human hippocampus, Nat. Neurosci. 26 (2023) 1994–2004; doi: 10.1038/s41593-023-01456-8

[7] A. Anta1, M. Bikson, A. Datta, B. Lafon, P. Dechent, L. C. Parra, W. Paulus, *Neuroimage* **85** (2014) 1040-1047; doi: 10.1016/j.neuroimage.2012.10.026

[8] J. Li, K.-M. Lee, K. Bai, Analytical and Experimental Investigation of Temporal Interference for Selective Neuromuscular Activation, *IEEE Trans. Neural Syst. Rehab. Eng*. **28** (2020) 3100-3112.

[9] K. Iszak, S. M. Gronemann, S. Meyer, A. Hunold, J. Zschüntzsch, M. Bähr, W. Paulus, A. Antal, Why Temporal Inference Stimulation May Fail in the Human Brain: A Pilot Research Study, *Biomedicines* **11** (2023) 1813; doi: 10.3390/ biomedicines11071813

[10] G.A. Korn, T.M. Korn, Mathematical Handbook for Scientists and Engineers. Dover Publications (2000), New York.